\documentclass[]{spie}  

\usepackage[dvips]{graphicx}
\usepackage{amssymb}
\usepackage{times}
\usepackage{pstricks}

\def\ls{\mathrel{\raise0.27ex\hbox{$<$}\kern-0.70em 
\lower0.71ex\hbox{{$\scriptstyle \sim$}}}}
\def\gs{\mathrel{\raise0.27ex\hbox{$>$}\kern-0.70em 
\lower0.71ex\hbox{{$\scriptstyle \sim$}}}}
\def\arcsec{\hbox{$^{\prime\prime}$}}

\title{Overview of the SuperNova/Acceleration Probe (SNAP)}

\author{\hspace{-6pt}\parbox{17cm}{\center G.~Aldering\supit{a},
C.~Akerlof\supit{b},
R.~Amanullah\supit{c},
P.~Astier\supit{d}, E.~Barrelet\supit{d},
C.~Bebek\supit{a}, L.~Bergstr\"{o}m\supit{c}, \\
J.~Bercovitz\supit{a}, G.~Bernstein\supit{e}, M.~Bester\supit{f},
A.~Bonissent\supit{g}, C.~Bower\supit{h}, W.~Carithers\supit{a}, \\
E.~Commins\supit{f}, C.~Day\supit{a},
S.~Deustua\supit{i}, R.~DiGennaro\supit{a}, A.~Ealet\supit{g},
R.~Ellis\supit{j},
M.~Eriksson\supit{c}, \\ A.~Fruchter\supit{k},
J-F.~Genat\supit{d}, G.~Goldhaber\supit{f},
A.~Goobar\supit{c}, D.~Groom\supit{a},
S.~Harris\supit{f},
P.~Harvey\supit{f}, \\ H.~Heetderks\supit{f}, S.~Holland\supit{a},
D.~Huterer\supit{l}, A.~Karcher\supit{a},
A.~Kim\supit{a},
W.~Kolbe\supit{a}, B.~Krieger\supit{a}, \\ R.~Lafever\supit{a},
J.~Lamoureux\supit{a},
M.~Lampton\supit{f}, M.~Levi\supit{a},
D.~Levin\supit{b}, E.~Linder\supit{a},
S.~Loken\supit{a}, \\
R.~Malina\supit{m}, R.~Massey\supit{n},
T.~McKay\supit{b}, S.~McKee\supit{b}, R.~Miquel\supit{a},
E.~M\"{o}rtsell\supit{c}, N.~Mostek\supit{h}, \\ S.~Mufson\supit{h},
J.~Musser\supit{h},
P.~Nugent\supit{a}, H.~Oluseyi\supit{a}, R.~Pain\supit{d},
N.~Palaio\supit{a}, D.~Pankow\supit{f}, \\ S.~Perlmutter\supit{a},
R.~Pratt\supit{f}, E.~Prieto\supit{m}, A.~Refregier\supit{n},
J.~Rhodes\supit{o},
K.~Robinson\supit{a}, N.~Roe\supit{a}, \\ M.~Sholl\supit{f},
M.~Schubnell\supit{b},
G.~Smadja\supit{p}, G.~Smoot\supit{f}, A.~Spadafora\supit{a},
G.~Tarle\supit{b}, \\ A.~Tomasch\supit{b}, H.~von der Lippe\supit{a},
D.~Vincent,\supit{d} 
J.~Walder\supit{a}, and G.~Wang\supit{a}}
\skiplinehalf
\supit{a}Lawrence Berkeley National Laboratory, Berkeley CA, USA\\
\supit{b}University of Michigan, Ann Arbor MI, USA\\
\supit{c}University of Stockholm, Stockholm, Sweden\\
\supit{d}CNRS/IN2P3/LPNHE, Paris, France\\
\supit{e}University of Pennsylvania, Philadelphia PA, USA\\
\supit{f}University of California, Berkeley CA, USA\\
\supit{g}CNRS/IN2P3/CPPM, Marseille, France\\
\supit{h}Indiana University, Bloomington IN, USA\\
\supit{i}American Astronomical Society, Washington DC, USA\\
\supit{j}California Institute of Technology, Pasedena CA, USA\\
\supit{k}Space Telescope Science Institute, Baltimore MD, USA\\
\supit{l}Case Western Reserve University, Cleveland OH, USA\\
\supit{m}CNRS/INSU/LAM, Marseille, France\\
\supit{n}Cambridge University, Cambridge, UK\\
\supit{o}Goddard Space Flight Center, Greenbelt MD, USA\\
\supit{p}CNRS/IN2P3/IPNL, Lyon, France
}

\authorinfo{Correspondence: e-mail galdering@lbl.gov; telephone 510-495-2203}

\begin{document} 
  \maketitle

\begin{abstract}

The SuperNova / Acceleration Probe (SNAP) is a space-based experiment
to measure the expansion history of the Universe and study both
its dark energy and the dark matter.  The experiment is motivated
by the startling discovery that the expansion of the Universe
is accelerating.  A 0.7~square-degree imager comprised of 36 large
format fully-depleted $n$-type CCD's sharing a focal plane with
36 HgCdTe detectors forms the heart of SNAP, allowing discovery and
lightcurve measurements simultaneously for many supernovae. The imager
and a high-efficiency low-resolution integral field spectrograph are
coupled to a 2-m three mirror anastigmat wide-field telescope, which
will be placed in a high-earth orbit.  The SNAP mission can obtain
high-signal-to-noise calibrated light-curves and spectra
for over 2000 Type Ia supernovae at redshifts between $z=0.1$ and 1.7.
The resulting data set can not only determine the amount of dark energy
with high precision, but test the nature of the dark energy by
examining its equation of state. In particular, dark energy due to a
cosmological constant can be differentiated from alternatives such as
``quintessence'', by measuring the dark energy's equation of state to
an accuracy of $\pm 0.05$, and by studying its time dependence.

\end{abstract}
%


\section{INTRODUCTION}
\label{sect:intro}  

In the past decade the study of cosmology has taken its first major
steps as a precise empirical science, combining concepts and tools from
astrophysics and particle physics.  The most recent of these results
have already brought surprises.   The Universe's expansion is
apparently accelerating rather than decelerating as expected solely due
to gravity.  This implies that the simplest model for the Universe ---
flat and dominated by matter --- appears not to be true, and that our
current fundamental physics understanding of particles, forces, and
fields is likely to be incomplete.

\begin{figure}[h]
\begin{center}
\includegraphics[height=4.5truein]{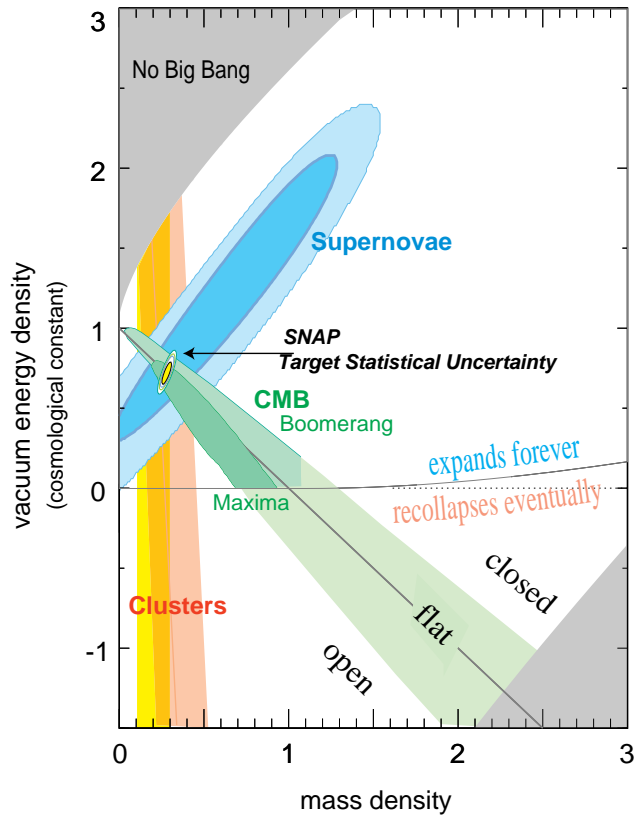}
\caption[$\Omega_M$---$\Omega_\Lambda$ confidence regions with current
SN, galaxy cluster, and CMB results.] {There is strong evidence for the
existence of a cosmological vacuum energy density.  Plotted are
$\Omega_M$---$\Omega_\Lambda$ confidence regions for current SN
\cite{p99}, galaxy cluster, and CMB results.  These results rule
out a simple flat, [$\Omega_M=1$, $\Omega_\Lambda=0$] cosmology.  Their
consistent overlap is a strong indicator for dark energy.  Also shown
is the expected confidence region from the SNAP satellite for an
$\Omega_M=0.28$ flat Universe.
}
\label{confcmbclust}
\end{center}
\end{figure}

The clearest evidence for this surprising conclusion comes from the
recent supernova measurements of changes in the Universe's expansion
rate that directly show the acceleration.  Figure~\ref{confcmbclust}
shows the results of Ref.~\citenum{p99} (see also Ref.~\citenum{riess98})
which compare the standardized brightnesses of 42 high-redshift Type~Ia
supernovae (SNe~Ia)
($0.18<z<0.83$) with 18 low-redshift SNe~Ia to find that for a flat
universe $\Omega_\Lambda=0.72\pm0.08$ ($\Omega_M=1-\Omega_\Lambda$),
or a deceleration parameter $q_0=-0.58$, and constrain the combination
$0.8\,\Omega_M-0.6\,\Omega_\Lambda$ to $-0.2\pm0.1$.

This evidence for a negative-pressure vacuum energy density is in
remarkable concordance with combined galaxy cluster measurements
\cite{bahcall00}, which are sensitive to $\Omega_M$, and CMB results
\cite{balbi2000,lange2001}, which are sensitive to the curvature
$\Omega_k$ (see Fig.~\ref{confcmbclust}).  Two of these three
independent measurements and standard inflation would have to be in
error to make the cosmological constant (or other negative pressure
dark energy) unnecessary in the cosmological models.

\begin{figure}[ht]
\begin{center}
\includegraphics[height=4.0truein]{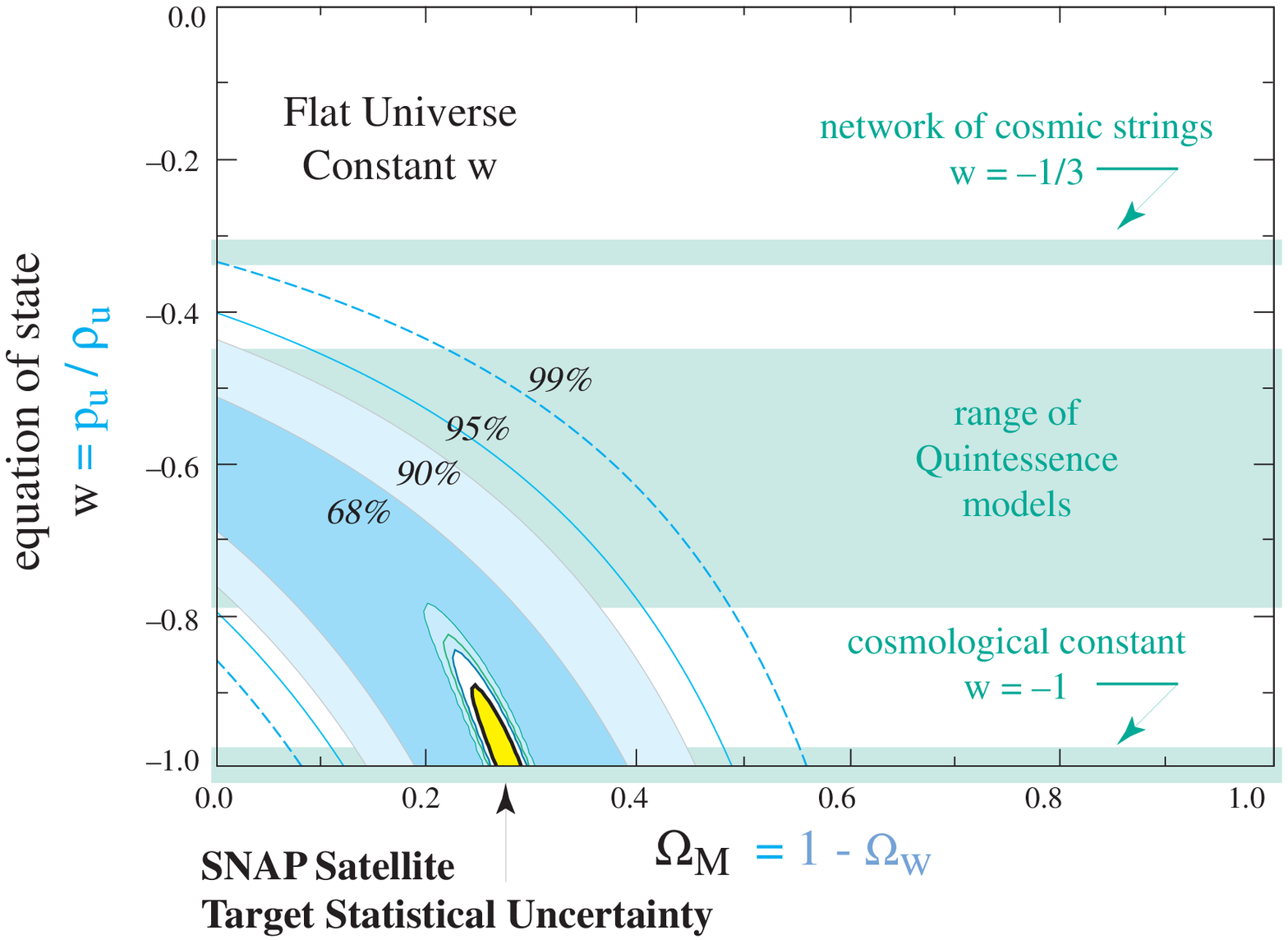}
\caption[Confidence regions in the $\Omega_{\rm M}$--$w$ plane.]
{Best-fit 68\%, 90\%, 95\%, and 99\% confidence regions in the
$\Omega_{\rm M}$--$w$ plane for an additional energy density component,
$\Omega_w$, characterized by an equation-of-state $w = p/\rho$.  (For
Einstein's cosmological constant, $\Lambda$, $w = -1$.) The fit is
constrained to a flat cosmology ($\Omega_{\rm M} + \Omega_w =1$).  Also
shown is the expected confidence region allowed by SNAP assuming $w=-1$
and $\Omega_M=0.28$.
}
\label{wconf}
\end{center}
\end{figure}

These measurements indicate the presence of a new, unknown energy
component that can cause acceleration, hence having equation of state
$w\equiv p/\rho<-1/3$.  This might be the cosmological constant.
Alternatively, it could be that this dark energy is due to some other
primordial field for which $\rho \ne -p$, leading to different
dynamical properties than a cosmological constant.  The fundamental
importance of a universal vacuum energy has sparked a flurry of
activity in theoretical physics with several classes of models being
proposed (e.g.  quintessence \cite{caldwell98,zlatev99},
Pseudo-Nambu-Goldstone Boson (PNGB) models \cite{frieman95,coble97},
cosmic defects \cite{vilenkin84,vilenkin94}).  Placing some constraints
on possible dark energy models, Refs.~\citenum{p99,garnavich98,ptw99}
find that for a flat Universe, the data are consistent with a
cosmological-constant equation of state with $0.2\ls\Omega_M\ls 0.4$
(Fig.~\ref{wconf}), or generally $w<-0.6$ at 95\% confidence level.
The cosmic string defect theory ($w=-1/3$) is already strongly
disfavored.

In this paper, we attempt to formulate a definitive supernova study
that will determine the values of the cosmological parameters and
measure the properties and test possible models for the dark energy.
In \S\ref{exec_uncertain} we identify and discuss how to minimize
systematic errors that fundamentally limit the precision with which
this probe can measure cosmological parameters.  A SN dataset
that maximizes the resolving power of the redshift-luminosity relation
under the constraint of these systematic errors is constructed in
\S\ref{asprobe}.  We present in \S\ref{proposed_experiment} the
SuperNova / Acceleration Probe (SNAP) whose observing strategy and
instrumentation suite is tailored to provide the data that satisfy both
our statistical and systematic requirements.

\section{CONTROL OF SYSTEMATIC UNCERTAINTIES}
\label{exec_uncertain}

Type Ia SNe have already proved to be an excellent distance indicator
for probing the dynamics of the Universe.  However, as we move toward
the  era of precision cosmology, we recognize that using the SN
redshift-luminosity distance relationship for measuring cosmological
parameters is fundamentally limited by potential systematic errors (as
are all cosmological probes).

\subsection{Known Sources of Systematic Errors}

Below are identified effects which any experiment that wishes to make
maximal use of this technique will need to recognize and control.
Following each item is the typical size of the effect on the SN
brightness, along with a rough estimate of the expected {\it
systematic} residual remaining after statistical correction for such
effects with the SNAP dataset.

\noindent
{\it Malmquist Bias:} A flux-limited sample preferentially detects
the intrinsically brighter members of any population of sources.
Directly correcting this bias would rely on knowledge of the SN Ia
luminosity function, which may change with lookback time.  A detection
threshold fainter than peak by at least five times the intrinsic SN~Ia
luminosity dispersion ensures sample completeness with respect to
intrinsic SN brightness, eliminating this bias ($\sim$5-10\%;0\%).

\noindent
{\it K-Correction and Cross-Filter Calibration:}  The current data set of
time and lightcurve-width-dependent SN spectra needed for K-corrections
is incomplete.  Judicious choice of filter sets, spectral time series
of representative SN~Ia, and cross-wavelength relative flux calibration
control this systematic ($\sim$0-10\%;$<0.5$\%).

\noindent
{\it Non-SN Ia Contamination:}  Observed supernovae must be positively
identified as SN~Ia.  As some Type~Ib and Ic SNe have spectra and
brightnesses that otherwise mimic those of SNe~Ia, a spectrum covering
the defining rest frame Si~II 6250\AA\ feature for every SN at
maximum will provide a pure sample ($\sim$10\%;0\%).

\noindent
{\it Galactic Extinction:} Supernova fields can be chosen toward the
low extinction Galactic poles. Future SIRTF observations will allow an
improved mapping between color excesses (e.g. of Galactic halo subdwarfs
in the SNAP field) and Galactic extinction by dust ($\sim$1-10\%;$<0.5$\%).

\noindent
{\it Gravitational Lensing by Clumped Mass:} Inhomogeneities along the
SN line of sight can gravitationally magnify or demagnify the
SN flux.  Since flux is conserved, the average of large numbers
of SNe per redshift bin will give the correct average brightness.
SNAP weak gravitational lensing measurements and micro-lensing studies
can further help distinguish whether or not the matter is in compact
objects ($\sim$1-10\%;$\sim0.5$\%).

\noindent
{\it Extinction by Extra-Galactic ``normal'' Dust:} Cross-wavelength
flux calibrated spectra will measure any wavelength dependent
absorption ($\sim$1-20\%;1\%).

\subsection{Possible Sources of Systematic Errors}

\subsubsection*{Extinction by Gray Dust} As opposed to normal dust,
gray dust is postulated to produce wavelength independent absorption in
optical bands.  Although physical gray dust grain models dim blue and
red optical light equally, the near-IR light ($\sim$1.2 $\mu$m) is less
affected.  Cross-wavelength calibrated spectra extending to wavelength
regions where ``gray'' dust is no longer gray will characterize the
hypothetical large-grain dust's absorption properties.  Armed with the
extinction -- color excess properties of the gray dust, broadband
near-infrared colors can provide ``gray'' dust extinction corrections
for SNe out to $z=0.5$.  Moreover, the gray dust will re-emit absorbed
starlight and thus contribute to the far-infrared background.  Deeper
SCUBA and SIRTF observations should tighten the constraints on the
amount of gray dust allowed.



\subsubsection*{Uncorrected Supernova Evolution}
Supernova behavior itself may have systematic variations depending,
e.g., on properties of its progenitor star or binary-star system.  The
distribution of these stellar properties is likely to change over
time---``evolve''---in a given galaxy, and over a population of
galaxies.  Nearby SNe~Ia drawn from a wide range of galactic
environments already provide an observed evolutionary range of
SNe~Ia\cite{hamuy96,hamuy00}.  The SNe differences that have been
identified in these data are well calibrated by the SN~Ia light curve
width-luminosity relation, leaving a 10\% instrinsic dispersion.  As of
yet, there is no evidence for systematic residuals after correction ---
aside from a few SNe with identified peculiarities --- although
observational errors would obscure such an effect.  It is not clear
whether any additional effects will be revealed with larger, more
precise and systematic, low-redshift SNe surveys\cite{aldering02}.

Theoretical models can identify observables that are expected to
display heterogeneity.  These key features, indicative of the
underlying initial conditions and physical mechanisms controlling the
SN, will be measured with SNAP, allowing statistical correction
for what would otherwise be a systematic error.  The state of empirical
understanding of these observables at the time SNAP flies will be
explicitly tested by SNAP measurements. Presently, we perform Fisher
matrix analyses on model spectra and lightcurves to estimate the
statistical measurement requirements, and ensure that we have
sensitivity to use subsamples to test for residual systematics at
better than the 2\% level. This approach reveals the main effects of
--- as well as covariance between --- the following observables:



\noindent
{\it Rise time from explosion to peak:} This is an
indicator of opacity, fused $^{56}$Ni mass and possible differences
in the $^{56}$Ni distribution. A 0.1~day uncertainty corresponds to
a 1\% brightness constraint at peak\cite{hoflich98}, and achieving
such accuracy requires discovery within $\sim$2 days of explosion,
on average, i.e. $\sim$30$\times$ fainter than peak.

\noindent
{\it Plateau level 45 days past peak:} The light curve plateau level
that begins $\sim$45 days past --- and more than 10$\times$ fainter than ---
peak is an important indicator of the
C/O ratio of the progenitor star, and fused $^{56}$Ni.  A 5\%
constraint on this plateau brightness corresponds to a 1\% constraint
on the peak brightness\cite{hoflich98}.  


\noindent
{\it Overall light curve timescale:} The ``stretch factor'' that
parameterizes the light curve time scale is affected by almost all
the aforementioned parameters since it tracks the SN~Ia's lightcurve
development from early to late times.  It is correlated with rise time
and plateau level, and it ties SNAP's controls for systematics to the
controls used in the current ground-based work.  A 0.5\% uncertainty in
the stretch factor measurement corresponds to a $\sim$1\% uncertainty
at peak \cite{p99}.

\noindent
{\it Spectral line velocities:} The velocities of several spectral
features throughout the UV and visible make an excellent diagnostic of
the overall kinetic energy of the SNe~Ia. Velocities
constrained to $\sim$250~km~s$^{-1}$ constrain the peak luminosity
$\sim$1\%\cite{hoflich98}, given a
typical SNe~Ia expansion velocity of 15,000~km~s$^{-1}$.

\noindent
{\it UV Spectral features:} The positions of various spectral features in
the restframe UV are strong indicators of the metallicity of the SNe~Ia.
By achieving a reasonable S/N on such features SNAP will be able to
constrain the metallicity of the progenitor to 0.1 dex\cite{lentz00}.
Spectral features in the restframe optical (Ca~II H\&K and Si~II at
6150 \AA) provide additional constraints on the opacity and luminosity
of the SN~Ia \cite{nugent95}.

By measuring all of the above features for each SN we can tightly
constrain the physical conditions of the explosion, making it possible
to recognize subsets of SNe with matching initial conditions.  The
approach which SNAP makes possible is to measure each feature well
enough to ensure a small luminosity range for each SNe subset
categorized according to physical condition. The expected residual
systematics from effects such as Malmquist bias, $K$-correction, etc,
discussed at the beginning of this section, total $\sim$2\%. Thus, a
subset analysis based on physical conditions should group SNe to within
$<$2\% in luminosity, and this is a goal in addition to a purely
statistical correction for any 2nd-parameter effects.

%

In addition to these features of the SNe themselves, we will
also study the host galaxy of the supernova.  We can measure the host
galaxy luminosity, colors, morphology, and the location of the SN
within the galaxy, even at redshifts $z\sim$1.7.  Such observations
are difficult or impossible from the ground.

\begin{figure}[ht]
\begin{center}
\includegraphics[height=4.5truein]{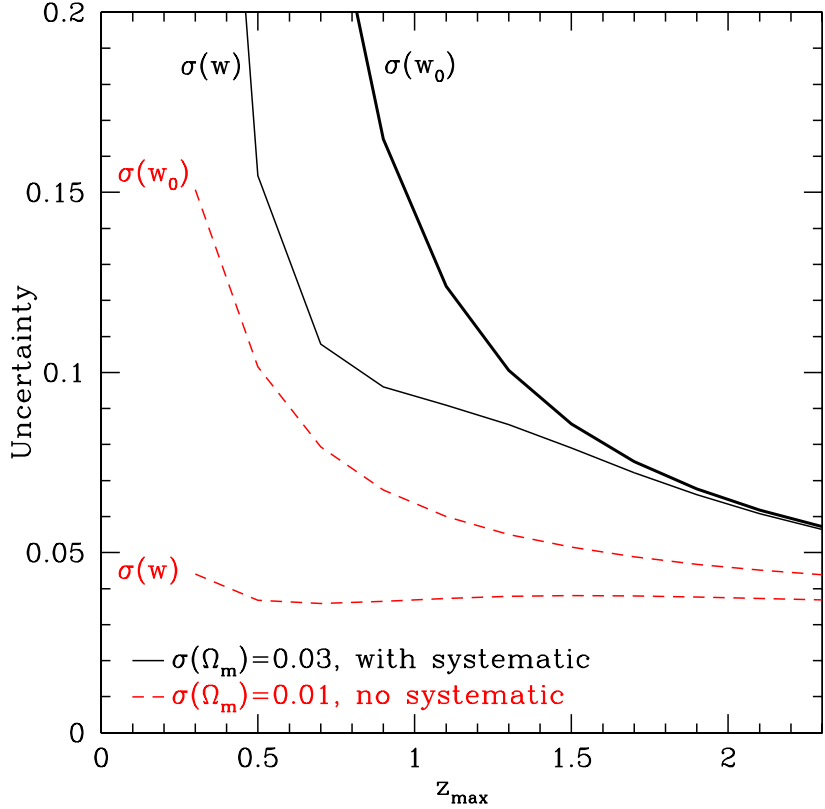}
\caption[Parameter estimation as a function of maximum redshift.]
{Accuracy in estimating the equation of state parameter, $w$, as a
function of maximum redshift probed in SN Ia surveys\cite{linder02}.  The
cases where $w$ is assumed constant in time are labeled as $\sigma(w)$,
while the cases where $w$ is allowed to vary with time as $w=w_0+w' z$
are labeled as $\sigma(w_0)$. The lower two curves assume that
the experiment is free of any systematic errors, while the upper two
curves are for the case where systematic errors are present at the 2\%
level. The top, heavy curve corresponds to the most realistic case.
It is clear that even with modest systematic errors good accuracy
requires probing to high redshift.  In all cases a flat universe is
assumed; a prior is also placed on $\Omega_M$, with a less constrained
prior of $\sigma_{\Omega_M}=0.03$ for the cases where systematics are
taken into account. \label{sci_deltaw.fig}}
\end{center}
\end{figure}

\section{SUPERNOVAE AS A PROBE OF THE DARK ENERGY}
\label{asprobe}
Our primary scientific objective is to use most efficiently the
leverage available in the redshift-luminosity distance relationship to
measure the matter and dark energy densities of the Universe with small
statistical and systematic errors, and also test the properties and
possible models for the dark energy.  We thus determine the number of
SNe we need to find, how they should be distributed in redshift,
and how precisely we need to determine each one's peak brightness.

The intrinsic peak-brightness dispersion of SNe~Ia after light-curve
shape and extinction correction is $\sim$10\%, so from a statistical
standpoint there is no need to measure the corrected peak brightness to
better than $\sim$10\%. With such statistical accuracy, a large sample
--- $\sim$2000 SNe~Ia --- is required to meet the measurement goals
given in Table~\ref{sci_errors.tab}. This large sample is also
necessary to allow model-independent checks for any residual
systematics or refined standardization parameters, since the sample
will have to be subdivided in a multidimensional parameter space of
redshift, lightcurve-width, host properties, etc.



The importance of using SNe~Ia over the full redshift range out to
$z\sim 1.7$ for measuring the cosmological parameters is demonstrated
in Fig.~\ref{sci_deltaw.fig} which shows the statistical uncertainty in
measuring the equation of state parameter, $w$, as a function of
maximum redshift probed in SN Ia surveys\cite{linder02}. This
simulation considers 2000 SNe~Ia in the range $0.1\le z\le z_{\rm max}$
measured with SNAP, along with 300 low-redshift SNe~Ia from the Nearby
Supernova Factory\cite{aldering02}.  A statistical error of 15\% was
assigned to each SN, which includes statistical measurement error and
intrinsic error.  A flat universe is assumed.  Two classes of models
were considered, one in which $w$ is forced to be constant with time
and a second in which $w$ varies with time according to the linear
expansion $w=w_0+w' z$.  The former case is applicable to a
cosmological constant or a network of topological defects, while the
latter case is applicable for all other models. For these two types of
models, two types of experiments were considered. The first is an
idealized experiment subject only to statistical errors and free of any
systematic errors. The second is a much more realistic model which
assumes both statistical and systematic errors, with the systematic
errors being very small like those SNAP can achieve. $\Omega_M$ is
constrained in both cases, with a prior which itself reflects the
impact of systematics, i.e., $\sigma_{\Omega_M}=0.01$ for an ideal
experiment and $\sigma_{\Omega_M}=0.03$ for a more realistic case.

From this figure we conclude that a SNe~Ia sample extending to redshifts
of $z>1$ is crucial for any realistic experiment in which there are
some systematic errors remaining after all statistical corrections are
applied. It is also clear that ignoring systematic errors can lead to
claims which are too optimistic.

Although current data indicate that an accelerating dark energy
density---perhaps the cosmological constant---has overtaken the
decelerating mass density, they do not tell us the actual magnitude of
either one.  These two density values are two of the fundamental
parameters that describe the constituents of our Universe, and
determine its geometry and destiny.  SNAP is designed to obtain
sufficient brightness-redshift data for a large enough range of
redshifts ($0.1<z<1.7$) that these absolute densities can each be
determined to unprecedented accuracy (see Fig.~\ref{confcmbclust}).
Taken together, the sum of these energy densities then provides a
measurement of the curvature of the Universe.  Assuming that the dark
energy is the cosmological constant, this experiment can simultaneously
determine mass density $\Omega_M$ to accuracy of $0.02$, cosmological
constant energy density $\Omega_{\Lambda}$ to $0.05$ and curvature
$\Omega_k=1-\Omega_M-\Omega_{\Lambda}$ to $0.06$.  The expected
parameter measurement precisions for this and other cosmological
scenarios  are summarized in Table~\ref{sci_errors.tab}; note that
these values are sensitive to the specific choice of dark energy model
and parameter priors.

\begin{table}
\caption{SNAP 1-$\sigma$ statistical and systematic uncertainties in parameter determination}
\begin{tabular}{cccccccccccc}
\hline\hline
 &\multicolumn{2}{c}{$\sigma_{\Omega_M}$} & & \multicolumn{2}{c}{$\sigma_{\Omega_{\Lambda}}$ (or $\sigma_{\Omega_{\rm D.E.}})$} & & \multicolumn{2}{c}{$\sigma_w$} & & \multicolumn{2}{c}{$\sigma_{w'}$} \\
 \cline{2-3} \cline{5-6} \cline{8-9} \cline{11-12} 
                       & stat & sys  & & stat  &   sys  & &stat  & sys& & stat & sys \\
\hline
 $w=-1$                &$0.02$&$0.02$& & $0.05$& $<0.01$& & ---  &  ---   &&  ---  & --- \\ 
 $w=-1$, flat          &  --- &  --- & & $0.01$& $0.02$ & & ---  &  ---   &&   --- & --- \\ 
 $w={\rm const}$, flat & ---  & ---  & & $0.02$& $0.02$ & &$0.05$& $<0.01$ &&  ---  & ---  \\ 
 $\Omega_M$, $\Omega_k$ known; $w={\rm const}$
                       & ---  &  --- & & ---   &  ---   & & 0.02 &  $<0.01$ & & ---   &  ---  \\ 
 $\Omega_M$, $\Omega_k$ known; $w(z)=w_0+w'\;z$ 
                       & ---  &  --- & & ---   &  ---   & & $0.08$& $<0.01$ & & $0.12$& $0.15$  \\
\hline
\label{sci_errors.tab}
\end{tabular}
\end{table}

The SNAP experiment is one of very few that can study the dark energy
directly, and test a cosmological constant against alternative dark
energy candidates.  Assuming a flat Universe with mass density
$\Omega_M$ and a dark energy component with a non-evolving equation of
state, this experiment will be able to measure the equation of state
ratio $w$ with accuracy of $0.05$ (for constant $w$), at least a factor
of five better than the best planned cosmological probes, including
systematic errors \cite{weller01,weller02}. With such a strong
constraint on $w$ we will be able to differentiate between the
cosmological constant and such theoretical alternatives as topological
defect models and a range of dynamical scalar-field (``quintessence'')
particle-physics models (see Fig.~\ref{wconf}).  Moreover, with data of
such high quality one can relax the assumption of the constant equation
of state, and test its variation with redshift, as predicted by many
theories including supergravity and M-theory inspired models.  These
determinations would directly shed light on high energy field theory
and physics of the early Universe.

CMB measurements from Planck will provide valuable complementarity and
cross comparison with SNAP, however CMB measurements are unable to
study $w(z)$. Other cosmological measurements are and will be
available, but those with sensitivity to $w(z)$ still have significant
systematics yet to be identified and overcome.  The simultaneous fits
of these measurements can improve constraints by as much as an order of
magnitude --- or they may not agree and upset our cosmological
understanding.

\begin{figure}[ht]
\begin{center}
\includegraphics[height=4.4truein]{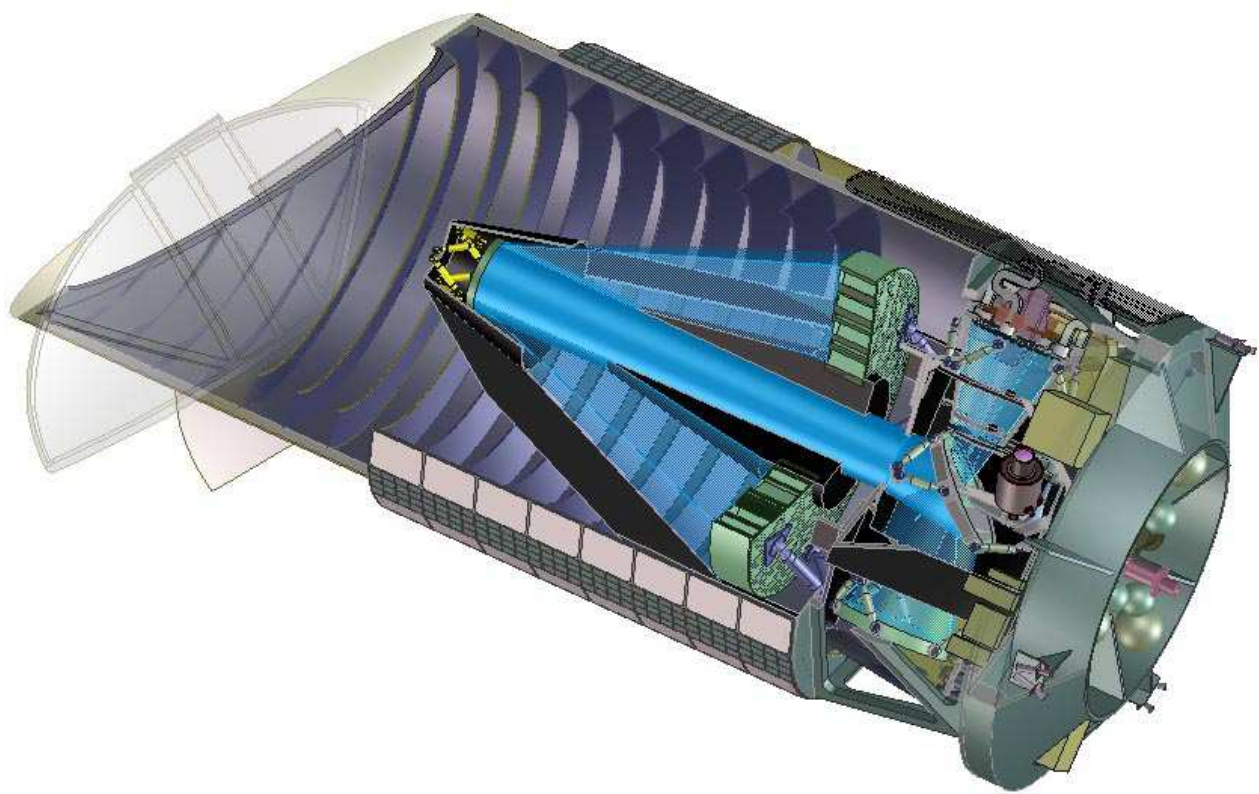}
\caption {A cross-sectional view of the SNAP satellite.  The principal
assembly components are the telescope, optical bench, instruments,
propulsion deck, bus, stray light baffles, thermal shielding and
entrance door.}
\label{snap_fig}
\end{center}
\end{figure}

\section{BASELINE EXPERIMENT}
\label{proposed_experiment}

To accomplish a rigorous test, discovery and study of more SNe and more
distant SNe (or any probe) is by itself insufficient.  As just shown,
we must address each of the systematic concerns while making precise SN
measurements, requiring a major leap forward in the measurement
techniques.  The science goals have thus driven us to the SNAP
satellite experiment that we describe in this section.

\subsection{Instrumentation}

\begin{figure}[htbp]
\begin{center}
\includegraphics[height=4.5truein]{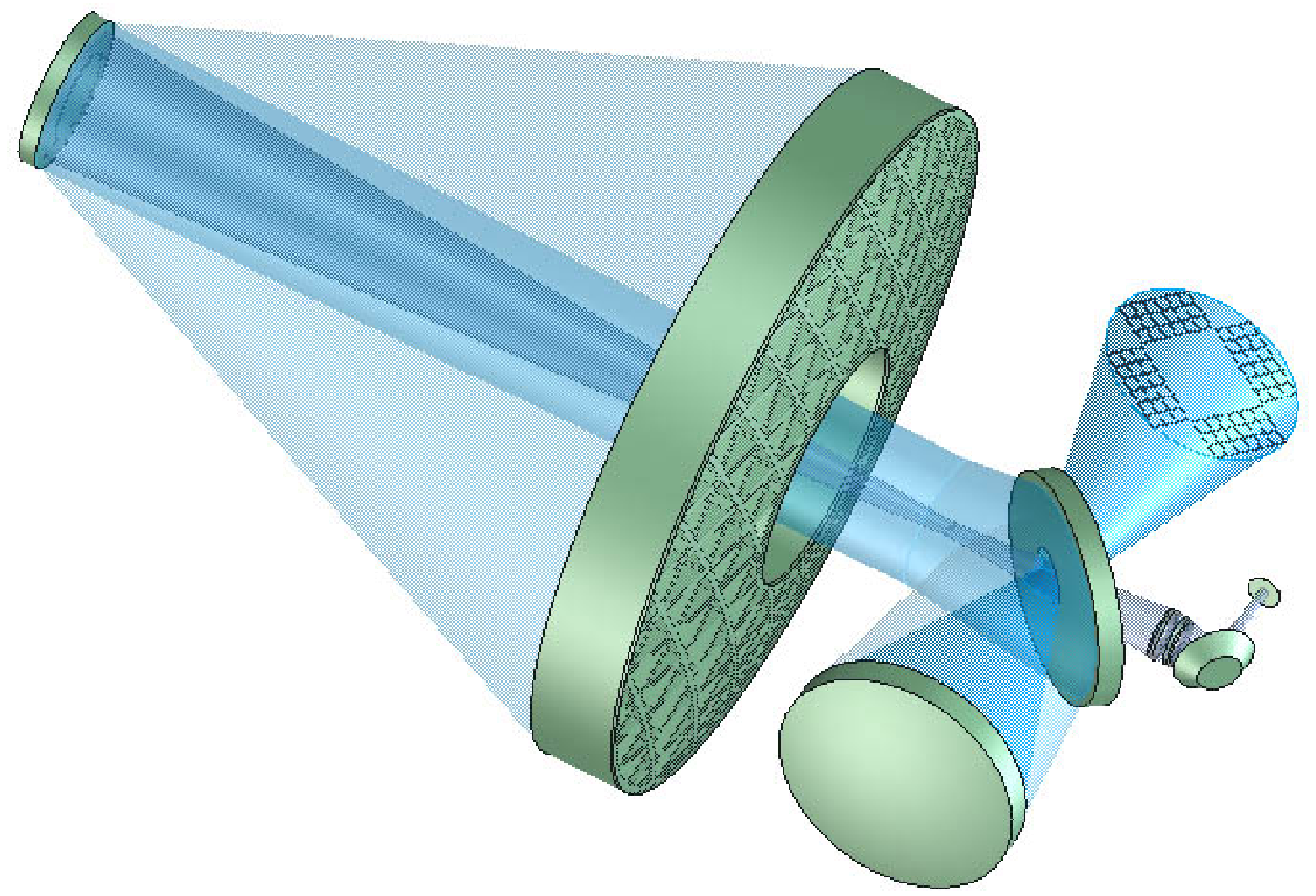}
\caption[Side view of our baseline optical configuration.] {Side view
of our baseline optical configuration, with a 2-m primary mirror,
a 0.45 meter secondary mirror, a folding flat, and a 0.7 meter tertiary
mirror. An optional auxiliary focus using the center of the field is possible.}
\label{side_view}
\end{center}
\end{figure}

The baseline for the SuperNova/Acceleration Probe is comprised of a
simple, dedicated combination of a 2-m telescope
three-mirror-anastigmat \cite{lampton02}, a 0.7~square-degree
optical--NIR imager and a low resolution (R$\sim$100) spectrograph
sensitive in the wavelength range 0.35 -- 1.7~$\mu$m.  A feedback loop
based on fast-readout chips on the focal plane is used to stabilize the
image.  A prototype of SNAP is illustrated in Fig.~\ref{snap_fig}.  The
mirror aperture is about as small as it can be before photometry and
spectroscopy at the requisite resolutions are no longer
zodiacal-light-noise limited.  A smaller mirror design would quickly
degrade the achievable S/N of the spectroscopy measurements, and
drastically reduce the number of SNe~Ia followed.  The
three-mirror-anastigmat, illustrated in Fig.~\ref{side_view}, achieves
a corrected field 1.4 degree in diameter, and the fraction of the focal
plane to populate with detectors has been chosen to obtain the
follow-up photometry of multiple SNe~Ia simultaneously.  A smaller
field would require multiple pointings of the telescope and again would
greatly reduce the number of SNe~Ia that could be followed.  The
spectrograph covers the wavelength range necessary to capture, over the
entire target redshift range, the Si~II 6150\AA\ feature that both
identifies SNe~Ia and provides a key measurement of the explosion
physics to probe the progenitor state.


The wide field of view of the SNAP imager allows simultaneous batch
discovery and photometry, and over the mission lifetime will yield
$\sim$2000 SNe with the proposed accuracy.  Even higher numbers of
more distant, less precisely measured SNe will be available in our data
set.  The wide-field imager covers 0.68~square-degrees of sky.
0.34~square~degrees consists of a mosaic of LBNL new technology $n$-type
high-resistivity CCD's \cite{holland99,stover00,groom00} that have high
($\sim$80\%) quantum efficiency for wavelengths between 0.35 and
1.0~$\mu$m.  Each of the 3.5k$\times$3.5k CCD's have 10.5 $\mu$m pixels
which give 0.1\arcsec\ per pixel with readout noise of 4e$^-$ and dark
current less than $0.002$~e$^-$~pixel$^{-1}$~sec$^{-1}$\cite{bebek02}.
Extensive radiation testing shows that these CCD's will suffer little
or no performance degradation over the lifetime of
SNAP\cite{bebek02b,bebek03}.  An additional 0.34~square~degrees is
covered by an array of 36 HgCdTe detectors\cite{tarle02}; we will use
commercially available 2k$\times$2k, $1.7~\mu$m cutoff devices with
18~$\mu$m pixels, high ($\sim60$\%) quantum efficiency, low
($\sim0.1$~e$^-$~pixel$^{-1}$~sec$^{-1}$) dark current, and 5~e$^-$
readout noise \cite{johnson00}. For exposures longer than a few minutes
the zodiacal light always dominates the detector noise.

\begin{figure}[htbp]
\begin{center}
\includegraphics[height=3.5truein]{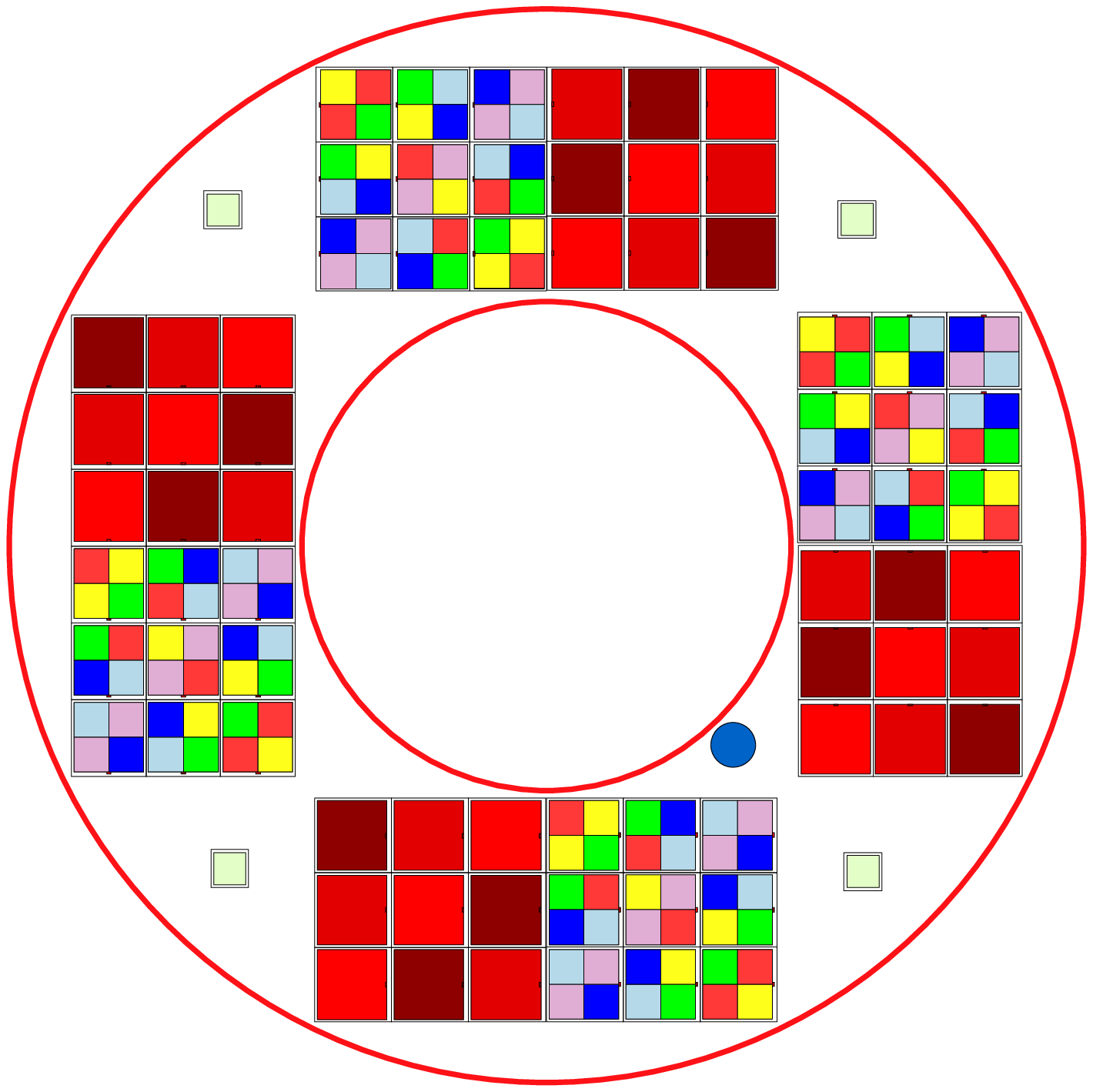}
\caption {The SNAP mosaic camera is tiled with 36 3.5k$\times$3.5k
high-resistivity CCD's and 36 HgCdTe detectors, covering
0.7~square degrees. The dectectors are arrayed to allow step
and stare sky coverage in orthogonal directions while coping with the
central obscuration that is necessary in a simple three-mirror anastigmat
telescope design. Each CCD is covered with four fixed filters, while each
HgCdTe has one fixed filter.}
\label{donut}
\end{center}
\end{figure}

Fixed filters are placed on each detector, arranged in the focal plane
such that each piece of sky can be observed in each filter with a shift
and stare mode of operation (Figure~\ref{donut}).  The relative areas
of each filter scale the cumulative exposure times to give limiting
fluxes per unit frequency that are nearly constant from 0.35--1.7~$\mu$m.
The imager will run a concurrent search and follow-up of SNe~Ia over
the entire redshift range $0.1\le z \le 1.7$ in wavelengths between 0.35
and 1.7~$\mu$m.

The SNAP spectrograph relies on an integral field unit (IFU) to
obtain an effective image of a 3\arcsec$\times3\arcsec$ field, split
into approximately 0.15\arcsec\ by 3\arcsec\ regions that are 
individually dispersed to obtain a flux at each position and wavelength
\cite{ealet02}.  A prism provides a high-throughput dispersive element
that makes possible observations of $z=1.7$ SNe~Ia with brightness 24
magnitudes at $\lambda=1.6~\mu$m (on the Vega system), with a 2-m aperture
telescope.  The broad SN spectral features require only low resolution
($R\sim100$). This resolution is achieved across the visible--NIR spectrum
with the prism used in single-pass for blue wavelengths and double-pass
for red wavelengths.  The visible detector is an LBNL CCD while the
NIR detector is HgCdTe, both as described above, but with a goal of improved
noise properties relative to the imager detectors. Together they provide
high quantum efficiency from $0.35 - 1.7\mu$m. In operation, the IFU
will allow simultaneous spectroscopy of a SN target and its surrounding
galactic environment; the 3\arcsec $\times$3\arcsec\ field of view also
removes any requirement for precise positioning of a SN target in a
traditional spectrograph slit and simplifies eventual subtraction of a
host galaxy spectrum.  This point is particularly important for absolute
flux calibration, because all of the SN light is collected with the IFU.
The spectrograph is thus designed to allow use of the spectra to obtain
photometry in any synthetic filter band that one chooses.

SNAP fits within a Delta-IV launch vehicle, and will be placed in
high-earth orbit in order to avoid thermal loading from the Earth. Passive
cooling of the focal plane is then achieved by restricting one side of
the spacecraft to face the Sun, and placing a radiator in the spacecraft
shadow. Passive cooling eliminates the need for cryogens, which would
otherwise increase the mass and decrease the lifetime of SNAP. Orbital
perigee will occur over the 11-m ground station at the Space Sciences
Lab (SSL) in Berkeley, allowing downloading of the $\sim3.5$~Tbits of
science data generated in each 3-day orbit. Mission operations will be
handled at SSL.

\subsection{Observation Strategy and Baseline Data Package}

This instrumentation will be used with a simple, predetermined
observing strategy designed to repeatedly monitor regions of sky near
the north and south ecliptic poles together covering 15 square degrees,
discovering and following SNe~Ia that explode in those regions.
Every field will be visited at least every four days, with sufficiently
long exposures that almost all SNe~Ia in the SNAP survey region will
be discovered within a few restframe days restframe days of explosion.
(SNe at much higher redshifts on average will be found slightly later
in their light curve rise times.) The periodic observation of fixed
fields ensures that every SN at $z < 1.7$ will be followed as it
brightens and fades.


This prearranged observing program will provide a uniform,
standardized, calibrated dataset for each SN, allowing for the
first time comprehensive comparisons across complete sets of SNe~Ia.
The following strategies and measurements will address, and often
eliminate, the statistical and systematic uncertainties described in
\S~\ref{exec_uncertain}.

\addtolength{\partopsep}{-2mm}
\begin{itemize}
\item Blind, batch-processed searching.
\item SNe~Ia at $0.1 \le z \le 1.7$.
\item Spectrum for every SN at maximum covering the rest
frame Si~II 6250\AA\ feature.
\item Spectral time series of representative SN~Ia, with cross-wavelength
relative flux calibration.
\item A light curve sampled at frequent, standardized intervals that
extends from $\sim$2-80 restframe days after explosion to obtain a
light-curve-width- and extinction-corrected peak rest-frame $B$ brightness to 10\%.
\item Multiple color measurements in 9 bands approximating
rest-frame $B$.
\item Final reference images and spectra to enable clean subtraction of 
host galaxy light. 
\end{itemize}
\addtolength{\partopsep}{2mm}

The quality of these measurements is as important as the 
time and wavelength coverage, so we require: 
\begin{itemize}
\item Control over S/N for these photometry and spectroscopy 
measurements, to permit the targeted high statistical significance 
for SNe over the entire range of redshifts.
\item Control over calibration for these photometry and spectroscopy 
measurements, with constant monitoring data collected to measure 
cross-instrument and cross-wavelength calibration.
\end{itemize}

Note that to date no single SN~Ia has ever been observed with this
complete set of measurements, either from the ground or in space, and
only a handful have a dataset that is comparably thorough.  With the
observing strategy described here, {\em every one} of $\sim$2000
followed SN~Ia will have this complete set of measurements.


Each systematic will either be measured, so that it can become part of
the statistical error budget, or bounded.  In addition the completeness
of the dataset will make it possible to monitor the physical properties
of each SN explosion, allowing studies of effects that have not
been previously identified or proposed.

Finally, we note that SNAP will be able to make complementary measurements
of the cosmological parameters using weak gravitational lensing and
Type~II supernovae. Moreover, the wide-field, deep, visible/NIR imaging
which SNAP will produce will be an amazing resource for many other areas
of study.

\section{CONCLUSION}

The surprising discoveries of recent years make this a fascinating new
era of empirical cosmology, addressing fundamental questions.  SNAP
presents a unique opportunity to extend this exciting work and advance
our understanding of the Universe.


\acknowledgements
This work was supported by the Director, Office of Science, of the
U.S. Department of Energy under contract No.  DE-AC03-76SF00098.


\bibliography{article}        

\begin{thebibliography}{10}

\bibitem{p99}
S.~{Perlmutter}, G.~{Aldering}, G.~{Goldhaber}, R.~A. {Knop}, P.~{Nugent},
  P.~G. {Castro}, S.~{Deustua}, S.~{Fabbro}, A.~{Goobar}, D.~E. {Groom}, I.~M.
  {Hook}, A.~G. {Kim}, M.~Y. {Kim}, J.~C. {Lee}, N.~J. {Nunes}, R.~{Pain},
  C.~R. {Pennypacker}, R.~{Quimby}, C.~{Lidman}, R.~S. {Ellis}, M.~{Irwin},
  R.~G. {McMahon}, P.~{Ruiz-Lapuente}, N.~{Walton}, B.~{Schaefer}, B.~J.
  {Boyle}, A.~V. {Filippenko}, T.~{Matheson}, A.~S. {Fruchter}, N.~{Panagia},
  H.~J.~M. {Newberg}, and W.~J. {Couch}, ``{Measurements of Omega and Lambda
  from 42 High-Redshift Supernovae},'' {\em Astrophys J.} {\bf 517},
  pp.~{565--586}, 1999.

\bibitem{riess98}
A.~G. {Riess}, A.~V. {Filippenko}, P.~{Challis}, A.~{Clocchiatti},
  A.~{Diercks}, P.~M. {Garnavich}, R.~L. {Gilliland}, C.~J. {Hogan}, S.~{Jha},
  R.~P. {Kirshner}, B.~{Leibundgut}, M.~M. {Phillips}, D.~{Reiss}, B.~P.
  {Schmidt}, R.~A. {Schommer}, R.~C. {Smith}, J.~{Spyromilio}, C.~{Stubbs},
  N.~B. {Suntzeff}, and J.~{Tonry}, ``{Observational Evidence from Supernovae
  for an Accelerating Universe and a Cosmological Constant},'' {\em Astron. J.}
  {\bf 116}, pp.~1009--1038, 1998.

\bibitem{bahcall00}
N.~A. {Bahcall}, J.~P. {Ostriker}, S.~{Perlmutter}, and P.~J. {Steinhardt},
  ``{The Cosmic Triangle: Revealing the State of the Universe},'' {\em Science}
  {\bf 284}, p.~1481, 1999.

\bibitem{balbi2000}
A.~{Balbi}, P.~{Ade}, J.~{Bock}, J.~{Borrill}, A.~{Boscaleri}, P.~{De
  Bernardis}, P.~G. {Ferreira}, S.~{Hanany}, V.~{Hristov}, A.~H. {Jaffe}, A.~T.
  {Lee}, S.~{Oh}, E.~{Pascale}, B.~{Rabii}, P.~L. {Richards}, G.~F. {Smoot},
  R.~{Stompor}, C.~D. {Winant}, and J.~H.~P. {Wu}, ``{Constraints on
  Cosmological Parameters from MAXIMA-1},'' {\em Astrophys. J.} {\bf 545},
  pp.~L1--LL4, 2000.

\bibitem{lange2001}
A.~E. {Lange}, P.~A. {Ade}, J.~J. {Bock}, J.~R. {Bond}, J.~{Borrill},
  A.~{Boscaleri}, K.~{Coble}, B.~P. {Crill}, P.~{de Bernardis}, P.~{Farese},
  P.~{Ferreira}, K.~{Ganga}, M.~{Giacometti}, E.~{Hivon}, V.~V. {Hristov},
  A.~{Iacoangeli}, A.~H. {Jaffe}, L.~{Martinis}, S.~{Masi}, P.~D. {Mauskopf},
  A.~{Melchiorri}, T.~{Montroy}, C.~B. {Netterfield}, E.~{Pascale},
  F.~{Piacentini}, D.~{Pogosyan}, S.~{Prunet}, S.~{Rao}, G.~{Romeo}, J.~E.
  {Ruhl}, F.~{Scaramuzzi}, and D.~{Sforna}, ``{Cosmological parameters from the
  first results of Boomerang},'' {\em Phys. Rev. D} {\bf 63}, p.~42001, 2001.

\bibitem{caldwell98}
R.~R. {Caldwell}, R.~{Dave}, and P.~J. {Steinhardt}, ``{Cosmological Imprint of
  an Energy Component with General Equation of State},'' {\em Physical Review
  Letters} {\bf 80}, pp.~1582--1585, 1998.

\bibitem{zlatev99}
I.~{Zlatev}, L.~{Wang}, and P.~J. {Steinhardt}, ``{Quintessence, Cosmic
  Coincidence, and the Cosmological Constant},'' {\em Physical Review Letters}
  {\bf 82}, pp.~896--899, 1999.

\bibitem{frieman95}
J.~A. {Frieman}, C.~T. {Hill}, A.~{Stebbins}, and I.~{Waga}, ``{Cosmology with
  Ultralight Pseudo Nambu-Goldstone Bosons},'' {\em Physical Review Letters}
  {\bf 75}, pp.~2077--2080, 1995.

\bibitem{coble97}
K.~{Coble}, S.~{Dodelson}, and J.~A. {Frieman}, ``{Dynamical {$\Lambda$} models
  of structure formation},'' {\em Phys. Rev. D} {\bf 55}, pp.~1851--1859, 1997.

\bibitem{vilenkin84}
A.~{Vilenkin}, ``{String-Dominated Universe},'' {\em Physical Review Letters}
  {\bf 53}, pp.~1016--1018, 1984.

\bibitem{vilenkin94}
A.~Vilenkin and E.~Shellard, {\em Cosmic Strings and other Topological
  Defects}, Cambridge University Press, Cambridge, U.K., 1994.

\bibitem{garnavich98}
P.~M. {Garnavich}, S.~{Jha}, P.~{Challis}, A.~{Clocchiatti}, A.~{Diercks},
  A.~V. {Filippenko}, R.~L. {Gilliland}, C.~J. {Hogan}, R.~P. {Kirshner},
  B.~{Leibundgut}, M.~M. {Phillips}, D.~{Reiss}, A.~G. {Riess}, B.~P.
  {Schmidt}, R.~A. {Schommer}, R.~C. {Smith}, J.~{Spyromilio}, C.~{Stubbs},
  N.~B. {Suntzeff}, J.~{Tonry}, and S.~M. {Carroll}, ``{Supernova Limits on the
  Cosmic Equation of State},'' {\em Astrophys. J.} {\bf 509}, pp.~74--79, 1998.

\bibitem{ptw99}
S.~{Perlmutter}, M.~S. {Turner}, and M.~{White}, ``{Constraining Dark Energy
  with Type Ia Supernovae and Large-Scale Structure},'' {\em Physical Review
  Letters} {\bf 83}, pp.~670--673, July 1999.

\bibitem{hamuy96}
M.~{Hamuy}, M.~M. {Phillips}, N.~B. {Suntzeff}, R.~A. {Schommer}, J.~{Maza},
  and R.~{Aviles}, ``{The Absolute Luminosities of the Calan/Tololo Type IA
  Supernovae},'' {\em Astron. J.} {\bf 112}, p.~2391, 1996.

\bibitem{hamuy00}
M.~{Hamuy}, S.~C. {Trager}, P.~A. {Pinto}, M.~M. {Phillips}, R.~A. {Schommer},
  V.~{Ivanov}, and N.~B. {Suntzeff}, ``{A Search for Environmental Effects on
  Type IA Supernovae},'' {\em Astron. J.} {\bf 120}, pp.~1479--1486, 2000.

\bibitem{aldering02}
G.~{Aldering} and {The Nearby Supernova Factory collaboration}, ``{An Overview
  of the Nearby Supernova Factory},'' in {\em Proc. SPIE Vol. 4836,
  Astronomical Telescopes and Instrumentation},   {\bf 4836}, 2003.

\bibitem{hoflich98}
P.~{Hoeflich}, J.~C. {Wheeler}, and F.~K. {Thielemann}, ``{Type IA Supernovae:
  Influence of the Initial Composition on the Nu cleosynthesis, Light Curves,
  and Spectra and Consequences for the Determination of Omega M and Lambda},''
  {\em Astrophys. J.} {\bf 495}, p.~617, 1998.

\bibitem{lentz00}
E.~J. {Lentz}, E.~{Baron}, D.~{Branch}, P.~H. {Hauschildt}, and P.~E. {Nugent},
  ``{Metallicity Effects in Non-LTE Model Atmospheres of Type IA Supernovae},''
  {\em Astrophys. J.} {\bf 530}, pp.~966--976, 2000.

\bibitem{nugent95}
P.~{Nugent}, M.~{Phillips}, E.~{Baron}, D.~{Branch}, and P.~{Hauschildt},
  ``{Evidence for a Spectroscopic Sequence among Type 1a Supernovae},'' {\em
  Astrophys. J.} {\bf 455}, p.~L147, 1995.

\bibitem{linder02}
E.~V. {Linder} and D.~{Huterer}, ``{Importance of SNe at $z>1.5$ to Probe Dark
  Energy},''

\bibitem{weller01}
J.~{Weller} and A.~{Albrecht}, ``{Opportunities for Future Supernova Studies of
  Cosmic Acceleration},'' {\em Physical Review Letters} {\bf 86},
  pp.~1939--1942, 2001.

\bibitem{weller02}
J.~{Weller} and A.~{Albrecht}, ``{Future supernovae observations as a probe of
  dark energy},'' {\em Phys. Rev. D} {\bf 65}, 2002.

\bibitem{lampton02}
M.~{Lampton} and {the SNAP Collaboration}, ``{SNAP Telescope},'' in {\em Proc.
  SPIE Vol. 4854, Astronomical Telescopes and Instrumentation},   {\bf 4854},
  p.~999, 2003.

\bibitem{holland99}
S.~E. {Holland}, M.~{Wei}, K.~{Ji}, W.~E. {Brown}, D.~K. {Gilmore}, R.~J.
  {Stover}, D.~E. {Groom}, M.~E. {Levi}, N.~{Palaio}, and S.~{Perlmutter},
  ``{Large Format CCD Image Sensors Fabricated on High Resistivity Silicon},''
  in {\em IEEE Workshop on Charge-Coupled Devices and Advanced Image Sensors},
  p.~239, 1999.

\bibitem{stover00}
R.~J. {Stover}, M.~{Wei}, K.~{Ji}, W.~E. {Brown}, D.~K. {Gilmore}, S.~E.
  {Holland}, D.~E. {Groom}, M.~E. {Levi}, N.~{Palaio}, and S.~{Perlmutter},
  ``{A 2Kx2K high resistivity CCD},'' in {\em Optical Detectors for Astronomy
  II: State-of-the-Art at the Turn of the Millenium},  p.~239, 2000.

\bibitem{groom00}
D.~E. {Groom}, S.~E. {Holland}, M.~E. {Levi}, N.~P. {Palaio}, S.~{Perlmutter},
  R.~J. {Stover}, and M.~{Wei}, ``{Back-illuminated, fully-depleted CCD image
  sensors for use in optical and near-IR astronomy},'' {\em Nuclear Instruments
  and Methods in Physics Research A} {\bf 442}, pp.~216--222, 2000.

\bibitem{bebek02}
C.~{Bebek} and {the SNAP Collaboration}, ``{SNAP Focal Plane},'' in {\em Proc.
  SPIE Vol. 4854, Astronomical Telescopes and Instrumentation},   {\bf 4854},
  2003.

\bibitem{bebek02b}
C.~{Bebek}, D.~{Groom}, S.~{Holland}, A.~{Karcher}, W.~{Kolbe}, J.~{Lee},
  M.~{Levi}, N.~{Palaio}, B.~{Turko}, M.~{Uslenghi}, M.~{Wagner}, and G.~{Wang}
  {\em IEEE Trans. Nucl. Sci.} .

\bibitem{bebek03}
C.~{Bebek}, D.~{Groom}, S.~{Holland}, A.~{Karcher}, W.~{Kolbe}, J.~{Lee},
  M.~{Levi}, N.~{Palaio}, B.~{Turko}, M.~{Uslenghi}, M.~{Wagner}, and
  G.~{Wang}, ``{Radiation testing},'' in {\em Proc. SPIE Vol. 4669},   {\bf
  4669}, 2003.

\bibitem{tarle02}
G.~{Tarle} and {the SNAP Collaboration}, ``{SNAP NIR detectors},'' in {\em
  Proc. SPIE Vol. 4850, Astronomical Telescopes and Instrumentation},   {\bf
  4850}, 2003.

\bibitem{johnson00}
J.~{Johnson}, E.~{Polidan}, A.~{Waczynski}, R.~{Hill}, G.~{Delo},
  M.~{Robberto}, C.~M. {Lisse}, and L.~{Cawley}, ``{Dark current measurements
  on a state of the art near-IR HgCdTe 1024$\times$1024 array},''

\bibitem{ealet02}
A.~{Ealet} and {the SNAP Collaboration}, ``{SNAP Focal Plane},'' in {\em Proc.
  SPIE Vol. 4854, Astronomical Telescopes and Instrumentation},   {\bf 4854},
  2003.

\end{thebibliography}
\bibliographystyle{spiebib}   

\end{document}